\documentclass{aa}  
\usepackage{color}
\usepackage{natbib,twoopt}
\usepackage[hyphenbreaks]{breakurl}
\usepackage[breaklinks]{hyperref}      
\bibpunct{(}{)}{;}{a}{}{,}             
\definecolor{cobalt}{rgb}{0.06, 0.2, 0.65}
\hypersetup{
  colorlinks,
  citecolor=cobalt,
  linkcolor=[rgb]{0.8, 0.2, 1.0},
  urlcolor=cobalt,
}
\makeatletter
  \newcommandtwoopt{\citeads}[3][][]{\href{http://adsabs.harvard.edu/abs/#3}%
    {\def\hyper@linkstart##1##2{}%
     \let\hyper@linkend\@empty\citealp[#1][#2]{#3}}}
  \newcommandtwoopt{\citepads}[3][][]{\href{http://adsabs.harvard.edu/abs/#3}%
    {\def\hyper@linkstart##1##2{}%
     \let\hyper@linkend\@empty\citep[#1][#2]{#3}}}
  \newcommandtwoopt{\citetads}[3][][]{\href{http://adsabs.harvard.edu/abs/#3}%
    {\def\hyper@linkstart##1##2{}%
     \let\hyper@linkend\@empty\citet[#1][#2]{#3}}}
  \newcommandtwoopt{\citeyearads}[3][][]%
    {\href{http://adsabs.harvard.edu/abs/#3}
    {\def\hyper@linkstart##1##2{}%
     \let\hyper@linkend\@empty\citeyear[#1][#2]{#3}}}
\makeatother
\usepackage{graphicx}
\usepackage{txfonts}
\usepackage{placeins}
\usepackage{titlesec}
\usepackage{midfloat}
\begin{document}

   \title{ADF22+: a declining faint end in the far-infrared luminosity function in the SSA22 protocluster at z=3.09}


   \author{Shuo Huang
          \inst{1,2}
          \and
          Hideki Umehata\inst{3,2}
          \and
          Ian Smail\inst{4}
          \and
        Kouichiro Nakanishi\inst{1,5}
        \and
        Bunyo Hatsukade\inst{1,5,6}
        \and
        Mariko Kubo\inst{7}
        \and
        Yoichi Tamura\inst{2}
        \and
        Tomoki Saito\inst{8}
        \and
        Soh Ikarashi\inst{9}
          }

   \institute{National Astronomical Observatory of Japan, 2-21-1 Osawa, Mitaka, Tokyo 181-8588, Japan
         \and
             Department of Physics, Graduate School of Science, Nagoya University, Furocho, Chikusa, Nagoya 464-8602, Japan
        \and
        Institute for Advanced Research, Nagoya University, Furocho, Chikusa, Nagoya 464-8602, Japan
        \and
        Centre for Extragalactic Astronomy, Department of Physics, Durham University, South Road, Durham DH1 3LE, UK
        \and
        Department of Astronomical Science, The Graduate University for Advanced Studies, SOKENDAI, 2-21-1 Osawa, Mitaka, Tokyo
181-8588, Japan
\and
Institute of Astronomy, Graduate School of Science, The University of Tokyo, 2-21-1 Osawa, Mitaka, Tokyo 181-0015, Japan
\and
Astronomical Institute, Tohoku University, 6-3, Aramaki, Aoba, Sendai, Miyagi, 980-8578, Japan
\and
Nishi-Harima Astronomical Observatory, Centre for Astronomy, University of Hyogo, 
407-2 Nishigaichi, Sayo, Sayo-gun, Hyogo 679-5313, Japan
\and
 Junior College, Fukuoka Institute of Technology, 3-30-1 Wajiro-higashi, Higashi-ku, Fukuoka, 811-0295, Japan 
        \\
             }

   \date{}
 
  \abstract{
  Protoclusters represent the densest regions of cosmic large-scale structure in the early universe and are the environment where present-day massive elliptical galaxies are assembled. Millimeter continuum emission offers a powerful probe of obscured star formation at high redshifts across various environments. In this paper, we present a deep ALMA 1.17 mm mosaic of the central 8 arcmin$^2$ ($\approx30$ comoving Mpc$^{2}$) region in the SSA22 protocluster at $z=3.09$ to study the faint dusty star-forming galaxy (DSFG) population. The continuum map achieves an RMS noise level of $\approx25$ $\mu$Jy beam$^{-1}$ at $\approx1\arcsec$ spatial resolution, $\approx2\times$ the depth of previous observation of this field. We detected 53 sources with a signal-to-noise ratio above 4.2, doubling the number of detections. Utilizing optical to mid-infrared ancillary data, we search for spectroscopic redshift and identify 18 of 53 as cluster members. For sources with more than two photometric data points in the near-infrared, stellar mass ($M_\star$) and star formation rate (SFR) from spectral energy distribution fitting are presented. The 1.17 mm number count shows $>2\times$ excess at flux density $\gtrsim1$ mJy but are consistent with blank field in fainter flux bins. The monochromatic far-infrared luminosity function of the SSA22 protocluster core region suggests a lack of faint DSFGs. All SSA22 protocluster member galaxies detected at 1.17 mm have SFR within the $M_\star$-SFR relation of general star-forming galaxies. Our results suggest that an early overdense environment like SSA22 protocluster predominantly enhances the formation of massive early-type galaxies in present-day galaxy clusters, but that the star formation in individual member galaxies is likely driven by gas supply along the cosmic web and occurs in a secular way.
  }


   \maketitle
%

\section{Introduction}
\par
Overdense regions in the high-redshift universe are thought to be the progenitors of present-day galaxy clusters and are referred to as protoclusters  \citep[e.g.,][]{2016A&ARv..24...14O}. In $\Lambda$CDM cosmology, structure formation occurs in a hierarchical manner. Peaks in the initial density fluctuation collapse into dark matter halos \citep{1972ApJ...176....1G} and define the filaments and sheets in the large-scale matter distribution \citep[the ``cosmic web'',][]{1996Natur.380..603B}. The most massive halos form from mergers of smaller ones at the intersecting nodes of cosmic filaments \cite[e.g.,][]{2009MNRAS.398.1150B}, where large amounts of infalling materials along the cosmic web are thought to fuel star formation and give rise to protocluters of galaxies in the early universe \citep{2019Sci...366...97U}. Observations of protoclusters suggest that galaxy evolution is accelerated in them compared to the field, signposted by elevated levels of star formation \citep[e.g.,][]{2003Natur.425..264S,2014A&A...570A..55D,2015ApJ...808L..33C,2018Natur.556..469M}, higher active galactic nuclei fractions \citep[AGN, e.g.,][]{2009MNRAS.400..299L,2009ApJ...691..687L,2013ApJ...765...87L,2022A&A...662A..54T}, and eventually, early emergence of massive quiescent galaxies \citep[e.g.,][]{2021ApJ...919....6K,2022ApJ...926...37M,2023ApJ...945L...9I}. The rapid growth and quenching of massive galaxies in protoclusters can explain old red giant elliptical galaxies in the center of present-day clusters \citep[e.g.,][]{2005ApJ...632..137N}.
\par
Most of the star formation in massive galaxies at high redshift is obscured by dust \citep[$>90\%$ for log$(M_\star/M_\odot)>10.5$, e.g.,][]{2017ApJ...850..208W,2020MNRAS.494.3828D}, thus one of the most efficient ways to probe their stellar mass build-up is to observe the rest-frame far-infrared emission which originates from dust heating by young massive stars. There are numerous studies targeting high-redshift overdensities using single-dish submillimeter/millimeter (submm/mm) telescopes equipped with bolometer cameras \citep[e.g.,][]{2000ApJ...542...27I,2005MNRAS.363.1398G,2009Natur.459...61T,2009ApJ...691..560C,2014MNRAS.440.3462U,2014A&A...570A..55D,2016MNRAS.461.1719C,2018A&A...620A.202A,2024MNRAS.529.2290S,2024arXiv240616637W}. These observations have consistently found abundant bright submm/mm sources indicating intense star formation enshrouded by dust in protoclusters. While single-dish surveys provide a wide field-of-view (FOV), which is needed to map protocluster fields, the coarse angular resolution limits the achievable sensitivity due to confusion noise and complicates the identification of counterparts. Interferometers offer angular resolution high enough to overcome the confusion limit of single-dish telescopes, but the small FOV necessitates mosaicking, so interferometric protocluster surveys with a significant area are scarce. Existing interferometric surveys of protoclusters \citep[e.g.,][]{2017ApJ...835...98U,2018ApJ...856...72O} are often too shallow to compare with deep blank field mosaics \citep[e.g.,][]{2020ApJ...897...91G,2024ApJS..275...36F}. As such, the effect of protocluster environment on the abundance of fainter and more representative populations of dusty star-forming galaxies remains poorly explored. Investigating the dust content of other populations in protoclusters, such as Ly$\alpha$ blobs \citep[e.g.,][]{2021ApJ...918...69U} and Lyman break galaxies also requires substantially deeper data.
\par
The Special Selected Area 22 (SSA22) protocluster at $z=3.09$ was first discovered as a peak in redshift distribution of Lyman break galaxies \citep{1998ApJ...492..428S}, and then associated with a $>100$ Mpc scale structure identified by overdensity of Ly$\alpha$ emitters \citep{2004AJ....128.2073H,2012AJ....143...79Y} distributed along at least three filaments \citep{2005ApJ...634L.125M}. The large-scale filaments intersect at the core region of the SSA22 protocluster, where numerous massive galaxies \citep{2013ApJ...778..170K,2015ApJ...799...38K,2016MNRAS.455.3333K} and AGNs \citep{2009MNRAS.400..299L,2009ApJ...691..687L} are seen.
Since its discovery, it has been extensively studied from X-ray to radio wavelengths \citep[e.g.,][]{2004ApJ...606...85C,2004AJ....128.2073H,2009ApJ...691..687L,2009ApJ...692.1561W,2016MNRAS.460.3861K,2017ApJ...850..178A}.
Single-dish submm/mm surveys of the SSA22 field \citep{2005MNRAS.363.1398G,2009Natur.459...61T,2014MNRAS.440.3462U} have found substantial star formation activities obscured by dust. \citet{2017ApJ...835...98U} mapped the central 7 arcmin$^2$ ($\approx1$ Mpc$^2$) area at 1.1 mm using ALMA and revealed an increased number density of bright dusty star-forming galaxies (DSFGs) in the core region. The distribution of $z=3.09$ member DSFGs and AGNs overlaps with diffuse gas filaments seen in Ly$\alpha$ emission \citep{2019Sci...366...97U}, suggesting galaxy growth fueled by gas supply from the cosmic web.
However, the abundance of fainter DSFGs was not clear because of the small number of detections at an RMS sensitivity of 60 $\mu$Jy beam$^{-1}$ (equivalent to a $4\sigma$ SFR limit of $\approx50$ $M_\odot$ yr$^{-1}$ at $z=3.09$).
\par
In this paper, we extend the previous 1.1 mm continuum mosaic in the SSA22 core region \citep[ALMA Deep Field in SSA22, or ADF22+, see also][]{2017ApJ...835...98U} to $>2\times$ fainter flux limit to study the faint dusty star-forming galaxy population in this protocluster region.
This paper is organized as follows. In section \ref{section:obs} we describe the observations and data analysis. The results are presented in section \ref{section:res} and discussed in section \ref{section:giron}. We summarize our findings in section \ref{section:con}. 
Throughout this paper, we adopt Planck 2018 cosmological parameters: $H_0=67.4$ km s$^{-1}$ Mpc$^{-1}$, $\Omega_{m}=0.315$ and $\Omega_{\mathrm{\Lambda}}=0.685$ \citep{2020A&A...641A...1P}. At $z=3.09$, $1\arcsec$ corresponds to 7.8 physical kpc.
\begin{figure*}[!htp]
    \centering
    \includegraphics[width=0.7\linewidth,angle=270]{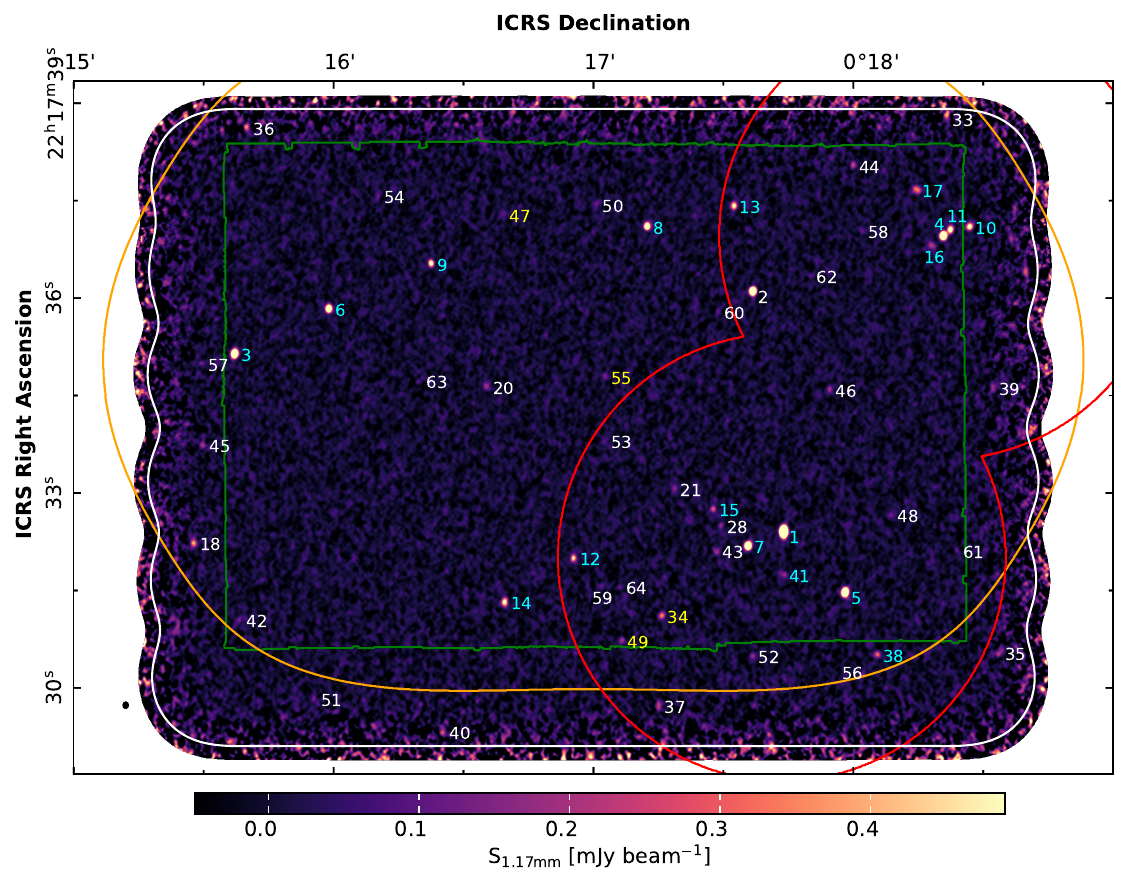}
    \caption{Left: 1.17mm continuum map displayed in flux range -2$\sigma$\textendash20$\sigma$. The black ellipse in the lower left indicates the beam FWHM. Sources with S/N$>4.2$ are indicated with their ID number in Table 1. The IDs of $z=3.09$ protocluster members confirmed by spectroscopy are highlighted with cyan color. Sources with non-zero values at $z=3.09$ in their photometric redshift probability functions (A34, A47, A49 and A55) are indicated with yellow color. The white contour indicates a primary beam response of 20\%. The colored contours show the areas with spectroscopic coverage from VLT/MUSE (green), \citet{2019Sci...366...97U} shallow CO mosaic (orange) and deep CO observations of A1 and A4 (red).}
    \label{fig1}
\end{figure*}
\begin{figure*}[!htp]
    \centering
    \includegraphics[width=0.9\linewidth]{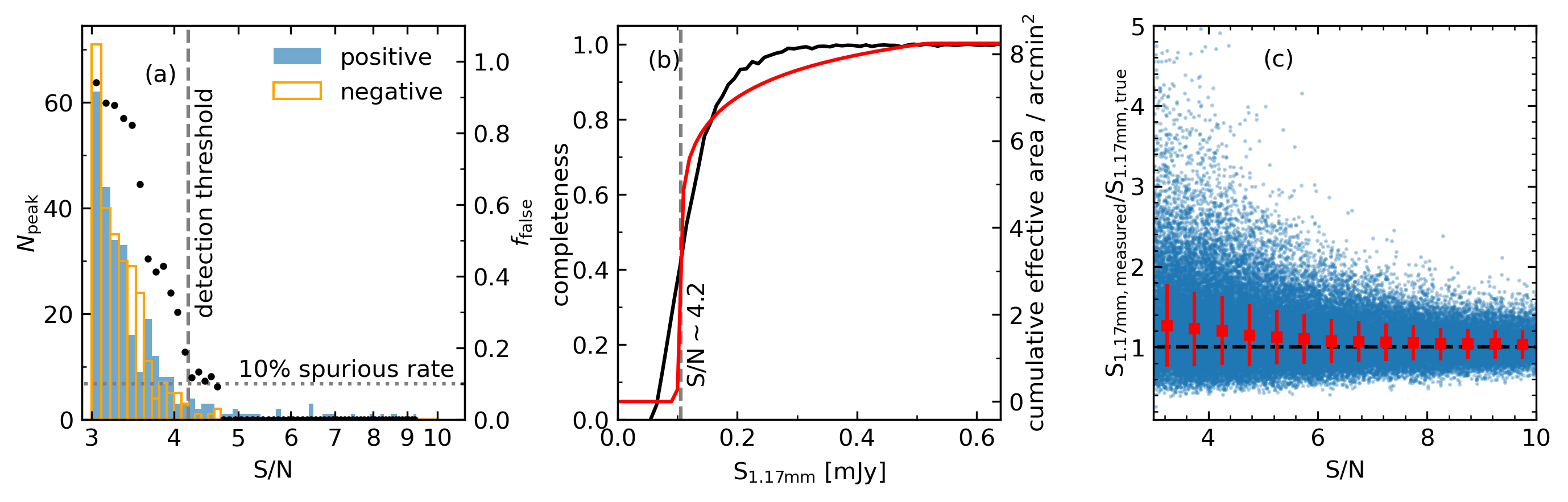}
    \caption{(a) Histograms show the number of positive and negative peaks as a function of S/N. The scatter plot indicates the false detection rate as a function of S/N. The dashed line marks the detection threshold of S/N$>4.2$, corresponding to a $12\%$ spurious rate (the dotted line). respectively. (b) completeness (black line) as a function of intrinsic flux and effective survey area (red line) in which a source can be detected with S/N$>4.2$ as a function of \textit{peak} flux per beam. The vertical line marks the typical flux at the detection threshold of S/N$>4.2$. (c) the ratio between measured and intrinsic fluxes as a function of S/N. Red squares with error bars indicate the median and standard deviation. The flux boosting is less than $\sim10\%$ at S/N$>4.2$.}
    \label{fig2}
\end{figure*}
\section{Observation and ancillary data}\label{section:obs}
\subsection{ALMA band 6 continuum mosaic}
The central $2\farcm5\times3\farcm3$ region of the SSA22 protocluster was mapped with the band 6 receiver in two ALMA projects from 2014 to 2018 (project codes \#2013.1.00162.S and \#2017.1.01332.S, PI: Umehata). The 103-pointing mosaic constructs a rectangular field centered on RA=$22^{\rm h}17^{\rm m}34^{\rm s}$ Dec=$+00\degr17\arcmin00\arcsec$ (Figure 1). The 2014 and 2015 observations were carried out in C-1 (angular resolution $\approx0\farcs8$, maximum recoverable scale $\approx7\farcs5$) and C-4 (angular resolution $\approx0\farcs4$, maximum recoverable scale $\approx3\farcs3$) array configurations with a single spectral tuning in sky frequency ranges of 252.98\textendash256.97 GHz and 268.98\textendash272.97 GHz. During the observations, J2148+069, Neptune, and Uranus were used as flux calibrators. Phase calibration was done by observing J2148+0657. The 2018 observations were conducted in C-1 (angular resolution $\approx0\farcs9$, maximum recoverable scale $\approx8\farcs3$) array configuration with three frequency tunings of 240.99\textendash244.97 GHz and 256.98\textendash260.97 GHz, 244.98\textendash248.97 GHz and 260.98\textendash264.97 GHz, 248.98\textendash252.97 GHz and 264.98\textendash268.97 GHz. These two projects provide continuous frequency coverage from 240.99 to 272.97GHz for the same field, covering the CO(9\textendash8) and H$_2$O (2$_{02}$\textendash1$_{11}$) lines at $z=3.09$. In 2018 observations, flux calibrators were J2148+0657, J2000-1748, and  J2253+1608. J2226+0052, J2156-0037, and	J2148+0657 were observed for phase calibration. The typical precipitable water vapor was 1.3 mm for 2014 and 2015 observations and 0.5\textendash2.3 mm for 2018 observations.
\par
The data were processed by the ALMA pipeline in the CASA \citep{2022PASP..134k4501C}  package. 
Imaging is performed using \texttt{tclean} task with Briggs weighting (robust=0.5), and \texttt{auto-multithresh} masking. We choose the \texttt{mtmfs} deconvolver with scale=[0] because the continuum sources are expected to be at most marginally resolved. The resulting 256.98 GHz (1.17 mm) continuum image is shown in Fig. 1. The synthesized beam has FWHM$=1\farcs3\times1\farcs1$ and position angle (PA)=$88.2\degr$. The RMS noise level is $\approx$25 $\mu$Jy beam$^{-1}$ before primary beam correction, $\approx2.4\times$ of the depth of $60\mu$Jy beam$^{-1}$ map from  \cite{2017ApJ...835...98U}.

\subsection{Ancillary data}
\subsubsection{ALMA band 3 spectroscopy}
The CO(3\textendash2) emission lines from two bright SSA22 protocluster member DSFGs with dense subgroups surrounding them, A1 and A4, were observed using the band 3 receiver (cycle 9 project \#2022.1.00680.S targeting A1 group and cycle 10 project \#2023.1.01206.S targeting A4 group, PI: Umehata). The cycle 9 program uses spectral tuning in sky frequency range 84.09\textendash87.80 GHz and 96.09\textendash99.80 GHz, while the cycle 10 program observes 84.06\textendash87.81 GHz and 96.09\textendash99.81 GHz, so that the lowest frequency spectral windows cover the CO(3\textendash2) line at $z=3.09$. The data are processed using the standard ALMA pipeline in the CASA package. Spectral cube is generated using the \texttt{tclean} task from pipeline-processed visibility data with natural weighting and 41.5 km s$^{-1}$ channel width, resulting in a median RMS noise level of 43 $\mu$Jy beam$^{-1}$ and FWHM$=1\farcs5\times1\farcs4$, PA=$69.4\degr$ synthesized beam for A1, and a median RMS noise level of 47 $\mu$Jy beam$^{-1}$ and FWHM$=3\farcs0\times2\farcs5$, PA=$87.5\degr$ synthesized beam for A4. In this paper, we solely use the data for redshift determination. Detailed analysis of the deep CO(3-2) data will be presented in future work. We also make use of the shallower CO(3-2) data in \citet{2019Sci...366...97U}, which covers a larger portion of the 1.17 mm continuum mosaic. 
\subsubsection{MUSE spectroscopy}
The Ly$\alpha$ emission line from the SSA22 protocluster was observed using the MUSE instrument on VLT UT4 (programs 099.A-0638 and 0101.A-0679, PI: Umehata). Two MUSE fields with 4.16 hours exposure per field are used to construct a $116\arcsec\times169\arcsec$ rectangular mosaic centered on RA=$22^{\rm h}17^{\rm m}34^{\rm s}.5$ Dec=$+00\degr17\arcmin00\arcsec$. The MUSE mosaic is almost co-spatial with the ALMA map, but the outer part of the ALMA map is not included. The cube covers the observed wavelength range 456 \textendash{ }935 nm with a spectral resolution of 0.125 nm. This spectral coverage includes Ly$\alpha$ (rest-frame 121.6 nm) at $z=3.09$. We refer the reader to \citet{2019Sci...366...97U} for details about the MUSE data, which also provides a compilation of source redshifts from ground-based optical and near-infrared spectroscopy.
\subsubsection{Optical and near-infrared imaging data}
The SSA22 field is imaged by various ground-based and space telescopes at optical and near-infrared (NIR) wavelengths. For ground-based imaging, we collected reduced data in CFHT/Megacam $u$ (K. Mawatari, private communication),  Subaru/SuprimeCam $BVRi^\prime z^\prime$ / NB359 / NB497 / NB816 / NB912 \citep{2004AJ....128.2073H,2009ApJ...692.1287I,2011MNRAS.412.2579N,2012AJ....143...79Y}, Subaru/HSC $grizY$ \citep{2018PASJ...70S...8A}, UKIRT/WFCAM $JK$ \citep{2007MNRAS.379.1599L} and Subaru/MOIRCS $Ks$ \citep{2021ApJ...919....6K} bands. Spitzer IRAC channel 1\textendash4 imaging data \citep{2009ApJ...692.1561W} are downloaded from the IRSA archive\footnote{\citet{SpitzerHeritageArchive}} and mosaicked using the \textsc{mopex} software. Part of the field is also imaged by the NIRCam onboard the James Webb Space Telescope (program GO-3547, PI: Umehata) in F115W, F200W, F356W, and F444W filters. 

\section{Analysis and results}\label{section:res}
\subsection{Source detection}
Source extraction is performed on the 1.17 mm continuum image before primary beam correction by using image segmentation routines in the \texttt{astropy} coordinated package \texttt{photutils} \citep{larry_bradley_2024_13989456}. We search for continuous pixel islands with amplitude above $0.5\times$RMS and total area larger than  $1\times$synthesized beam as initial source candidates.
The false detection rate in each peak signal-to-noise ratio (S/N) bin is calculated by repeating source extraction on reversed data and comparing the cumulative number of positive and negative peaks above a given S/N:
\begin{equation}
    f_{\rm false}({\rm S/N}>s) = \frac{N_{\rm negative}({\rm S/N}>s)}{N_{\rm positive}({\rm S/N}>s)}
\end{equation}
In Fig. 2a, we show the false detection rate as a function of peak S/N. Based on these results, we filter the initial source candidates by requiring peak S/N$>4.2$ and create the final source catalogue (Appendix A, see also Fig. 1) from 53 sources, corresponding to a spurious rate of $<12\%$. For each detection, we search for a counterpart in \citet{2017ApJ...835...98U} main and supplementary catalogues (A1\textendash A32) using a matching radius of $0.425\times$beam FWHM ($\approx0\farcs7$, equivalent to $1\sigma$ of Gaussian). If no match is found, a new ID starting from A33 is assigned in the order of descending flux. All 18 sources in the main catalogue of \citet{2017ApJ...835...98U} are detected in the new deeper mosaic, but only 3 out of 14 in the supplementary catalogue with S/N$=4.0$\textendash5.0 (or 2 of 8 sources with $4.2\leq$S/N$<5.0$) are recovered. Considering the poor confirmation of \citet{2017ApJ...835...98U} catalogue at S/N$<5$, we mark the new sources with $4.2<{\rm S/N}<5.0$ as tentative detection. 
\par
Source position and flux are measured using the CASA \texttt{imfit} task. The detection completeness is estimated by injecting point sources into random blank positions in the image before primary beam correction, and then running the same source detection script. The artificial source is considered recovered if it is detected at peak S/N$>4.2$ within $0\farcs7$ from the injected position. The completeness as a function of source flux is shown in Figure 2b. From the results, we also calculate the ratio between measured and true fluxes (Figure 2c). For sources with peak S/N$<20$, we divide their 1.17 mm fluxes from \texttt{imfit} task by the median measured-to-true flux ratio at the corresponding S/N to correct for flux boosting.
\subsection{Counterpart identification and source properties}
To search for counterparts at optical and near-infrared wavelengths, we first align the images to the ALMA coordinate frame and subtract the background. A gallery of image stamps at the ALMA source position is provided in Appendix C. The ground-based images have a homogenized point spread function (PSF) FWHM of $\approx1\farcs2$. Photometry on these images is performed in $2\arcsec$ diameter circular aperture centered on the ALMA position and corrected for aperture loss using the point spread function model in each band. Flux errors are estimated as the standard deviations of fluxes from 1000 randomly placed apertures at blank positions. Fluxes in Spitzer IRAC bands are measured from point source decomposition of the image \citep{2012ApJS..203...23H} in the same aperture as optical images, which allows the source centroid to wander by up to $\approx0\farcs8$. In Table \ref{tablea1}, sources with S/N$>3$ in any of the $Ks$, IRAC1 and IRAC2 bands (median $3\sigma$ depth = 0.3 $\mu$Jy, 0.3 $\mu$Jy and 0.4 $\mu$Jy) are marked as NIR detected. Three sources (A33, A58 and A64) have a nearby NIR source located  $>0\farcs7$ away from the ALMA position and no photometry. They are marked as potential lensed sources or no counterpart (L/N) in our catalogue.
\par
The majority of the brighter sources with an entry in \cite{2017ApJ...835...98U} have redshifts determined by previous studies. For all but one of these sources (A15), we collect their literature redshifts. For fainter sources and A15, we inspect their MUSE and ALMA band 3 spectra when available and search for redshift solutions. The redshifts and identified lines are presented in the source table (Appendix A). Details of the line identification are discussed in Appendix B. For sources without spectral line detected, we compute photometric redshift using the \textsc{eazy} code \citep{2008ApJ...686.1503B} if more than two data points in optical and NIR bands and report the resulting 16th-50th-84th percentile range in Table 1. Stellar mass and SFR are derived from optical/NIR and ALMA 1.17 mm fluxes with the SED fitting method in \citet{2024arXiv240609890T}. In Fig. 5, we plot spectroscopically confirmed SSA22 protocluster members in $M_\star$-SFR plane. For a comparison with coeval star-forming galaxy population in the same epoch, we also plot the so-called star-forming main sequence relation (SFMS) between $M_\star$ and SFR from \citet{2014ApJS..214...15S} and random field DSFGs at $2.8<z<3.4$ \citep[][Ikarashi et al. in prep]{2018A&A...620A.152F,2020A&A...643A..30F,2020MNRAS.494.3828D,2020MNRAS.495.3409S}.
\begin{figure}[htp]
    \centering
    \includegraphics[width=0.99\linewidth]{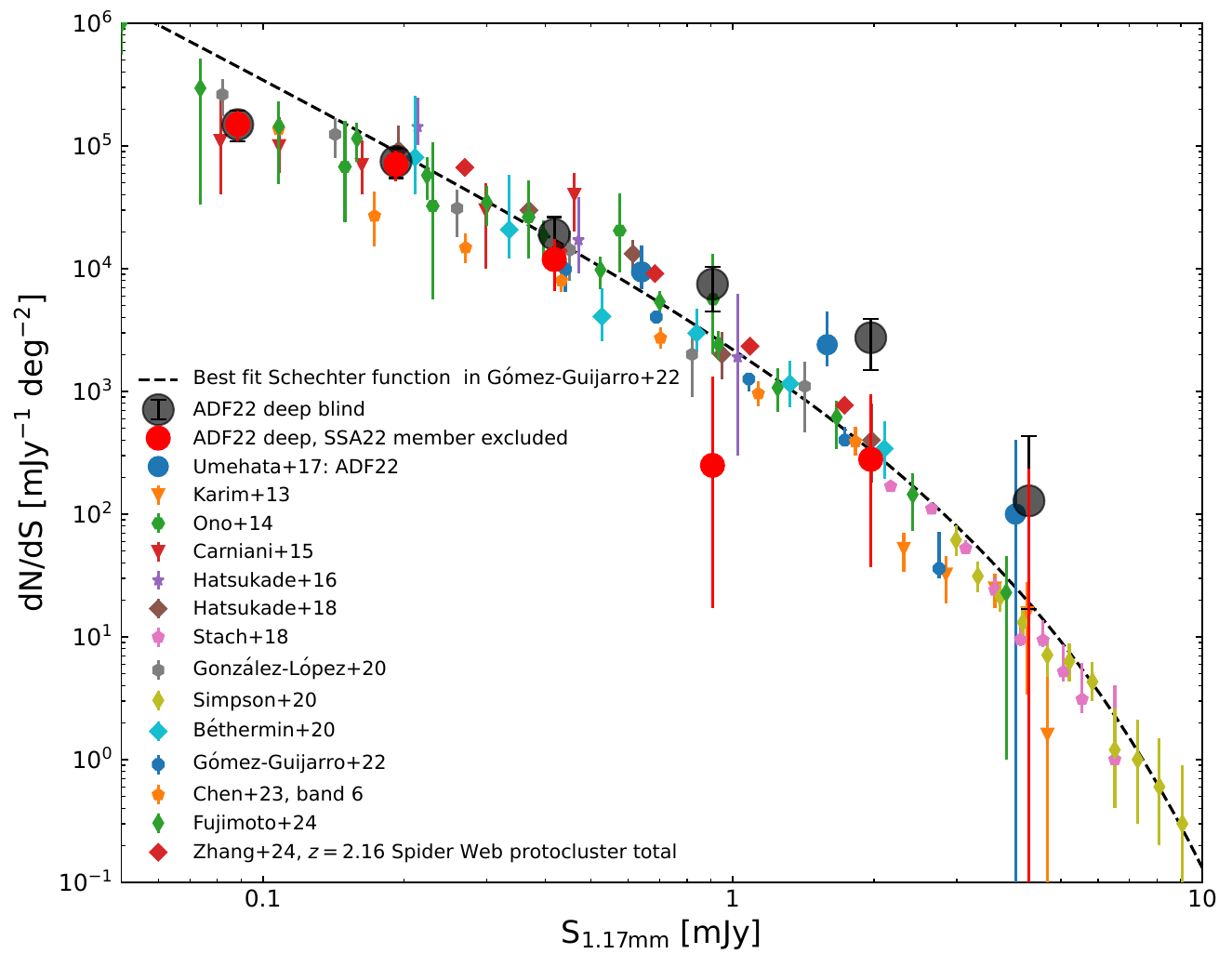}
    \caption{1.17 mm source number count of the deep SSA22 field. For comparison, we show interferometric number counts of random fields from literature, with fluxes converted to 1.17 mm: \citet{2013MNRAS.432....2K}, \citet{2014ApJ...795....5O}, \citet{2015A&A...584A..78C}, \citet{2016PASJ...68...36H}, \citet{2018PASJ...70..105H}, \cite{2018ApJ...860..161S}, \citet{2020ApJ...897...91G}, \citet{2020MNRAS.495.3409S}, \citet{2020A&A...643A...2B}, \citet{2022A&A...658A..43G}, 
    \citet{2023MNRAS.518.1378C},\cite{2024ApJS..275...36F}. We also present the total number count of the $z=2.16$ Spiderweb protocluster \citep{2024A&A...692A..22Z}. The dashed line shows the best-fit Schechter function from \citet{2022A&A...658A..43G}.}
    \label{fig3}
\end{figure}
\subsection{1.17 mm number count}
The blind differential number count in a flux bin with width $\Delta $S is calculated by summing the correction factor of each source in the bin:
\begin{equation}
    \frac{dN}{d\rm S}=\frac{1}{\Delta \rm S}\sum\frac{1-f_{\rm false}}{{\rm completeness}\times{\rm effective\: area}}.
\end{equation}
The effective area as a function of flux is defined as the area in the mosaic where a source can be detected with S/N$>4.2$ as shown in Figure 2b. We evaluate the number count from 0.06 to 6.3 mJy in six bins placed evenly in logarithmic space. Field-only number counts are also calculated by assigning a weight of zero to confirmed SSA22 members with spectroscopic redshift $z\approx3.09$. To estimate uncertainty, we resample the flux column of the source table (Appendix A) 1000 times, assuming the flux error follows a Gaussian distribution, and add the resulting standard deviation of the number count in quadrature to the Poisson error as the final uncertainty. The results are listed in Table 1 and plotted in Figure 4, together with a collection of literature results of blank fields \citep{2013MNRAS.432....2K,2014ApJ...795....5O,2015A&A...584A..78C,2016PASJ...68...36H,2018PASJ...70..105H,2018ApJ...860..161S,2020ApJ...897...91G,2020MNRAS.495.3409S,2020A&A...643A...2B,2022A&A...658A..43G,2023MNRAS.518.1378C,2024ApJS..275...36F} and protocluster field \citep{2024A&A...692A..22Z}. The literature fluxes are scaled to 1.17 mm by assuming a modified black body with dust temperature $=35$ K, emissivity $\beta=1.8$, and median redshift $z=2.5$ of the DSFG population \citep[e.g.,][]{2022A&A...658A..43G,2020MNRAS.494.3828D}. The resulting number count shows consistency with the field average for flux density $\lesssim1$ mJy but obvious excess at brighter fluxes.
\par
With a similar Monte Carlo method, we also compute the rest-frame 250 $\mu$m luminosity function $\phi$ for the core region of the SSA22 protocluster \citep[$3.08<z<3.10$,][]{2005ApJ...634L.125M}. Assuming a modified black body with dust temperature $=35$ K, emissivity $\beta=1.8$, and redshift $z=3.09$, the 1.17 mm (rest-frame 285 $\mu$m) luminosity is divided by 0.72 to convert to that at rest-frame 250 $\mu$m. Protocluster membership is based on spectroscopic redshifts or the probability of photometric redshifts (see the next section) falling into the range of the core region. We resample the catalogue 40,000 times and find the probability that one or more photometric redshift sources are protocluster members to be $\approx6\%$. However, we note that membership based on photometric redshift can be unreliable: the two sources (A34 and A49) with photometric redshift closest to 3.09 lie in the deep CO(3-2) footprint but are undetected while their continuum fluxes are comparable to A42 at $z=3.098$. As such, we mainly consider the results based on spectroscopic membership.
The monochromatic luminosity function is listed in Table 2 and plotted in Fig. 4.
\begin{table}
\caption{\label{table1}1.17 mm source number count.}
\centering
\begin{tabular}{ccll}
\hline\hline
 Flux bin &S$_{\rm 1.17mm}$ &\multicolumn{2}{c}{dN/dS}\\{}
 [mJy]&[mJy] &\multicolumn{2}{c}{[mJy$^{-1}$ deg$^{-2}$]}\\
\hline
&&blind&$z=3.09$ excluded\\
\hline
0.060\textendash0.130 &0.088 &$149000^{+43200}_{-41100}$&$149000^{+43300}_{-41200}$\\
0.130\textendash0.283 &0.192 &$74700^{+20900}_{-20000}$&$69900^{+20200}_{-19300}$\\
0.283\textendash0.615 &0.417 &$19000^{+7260}_{-6720}$&$11900^{+5590}_{-5320}$\\
0.615\textendash1.335 &0.906 &$7470^{+2860}_{-3010}$&$254^{+1060}_{-236}$\\
1.335\textendash2.901 &1.968 &$2740^{+1160}_{-1240}$&$279^{+660}_{-242}$\\
2.901\textendash6.300 &4.275 &$128^{+303}_{-111}$&$0^{+236}_{-0}$\\
\\
\hline
\end{tabular}
\end{table}
\begin{table}
\caption{\label{table2}Rest-frame 250 $\mu$m luminosity function of the SSA22 protocluster core region at $z=3.09\pm0.01$. The value in the faintest bin with photometric redshift membership included is quoted in parentheses.}
\centering
\begin{tabular}{ccc}
\hline\hline
 luminosity bin &luminosity  &$\phi$\\ 
 log($L$ [W Hz$^{-1}$])&log($L$ [W Hz$^{-1}$]) & log($\phi$ [Mpc$^{-3}$ dex$^{-1}$])\\
\hline
24.5\textendash24.9&24.7&$-2.79_{-0.90}^{+0.80}$($-2.72_{-0.89}^{+0.74}$)\\
24.9\textendash25.3& 25.1&$-1.57_{-0.32}^{+0.20}$\\
25.3\textendash25.7&25.5&$-1.51_{-0.26}^{+0.21}$\\
25.7\textendash26.1&25.9&$-1.76_{-0.35}^{+0.28}$\\
26.1\textendash26.5&26.3&$-2.36_{-0.88}^{+0.53}$\\
\hline
\end{tabular}
\end{table}
\begin{figure}[!ht]
    \centering
    \includegraphics[width=0.99\linewidth]{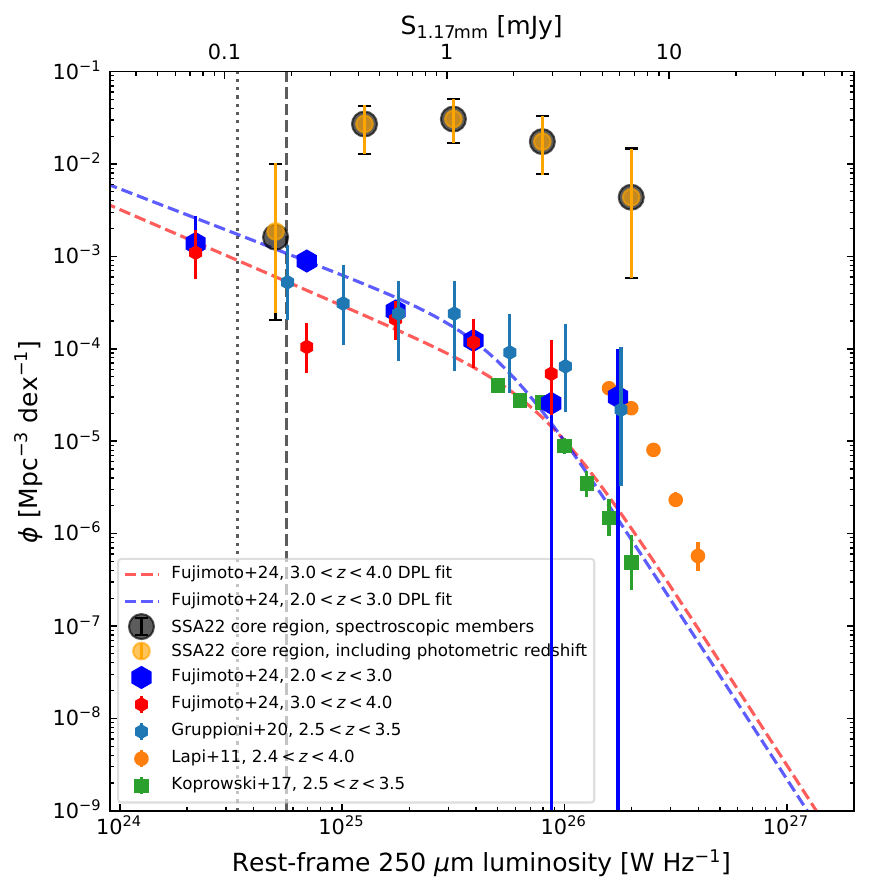}
    \caption{250 $\mu$m monochromatic luminosity function of the SSA22 protocluster core region at $z=3.09$. The blank field results at similar redshifts from single-dish \citep{2011ApJ...742...24L,2017MNRAS.471.4155K} or ALMA surveys \citep{2020A&A...643A...8G,2024ApJS..275...36F}, and double powerlaw (DPL) fits in \citep{2024ApJS..275...36F} are also plotted. The vertical dashed and dotted lines mark $90\%$ and $50\%$ completeness, respectively. Total IR luminosities are converted to that at 250 $\mu$m by assuming a modified blackbody with a temperature of 35 K and emissivity of 1.8.}
    \label{fig4}
\end{figure}

\section{Discussion}\label{section:giron} 

\subsection{overabundant bright dusty star-forming galaxies in protocluster environment}
From Fig. 3, it is clear that the SSA22 protocluster field shows an excess of sources with 1.17 mm flux density $\gtrsim1$ mJy as found by \cite{2017ApJ...835...98U}, but approaching the same number density as random fields at fainter fluxes. 
Adopting the best-fit Schechter function from \cite{2022A&A...658A..43G} as field average, the observed source surface density is $2.7^{+1.0}_{-1.1}$, $8.3^{+3.5}_{-3.7}$, $6.7^{+16.0}_{-5.9}$ times of that in blank field at 0.91, 2.0, and 4.2 mJy, while at 0.42 mJy, the source surface density is $1.1^{+0.4}_{-0.4}$ times of field average.
When known SSA22 members are excluded, the number count becomes roughly consistent with random fields, and the residual deviation at flux $\gtrsim1$ mJy can be attributed to the small number of bright field sources in the surveyed area. An excess of bright sources has also been observed in the $z=3.25$ MQN01 field \citep{2024A&A...684A.119P} and the Spiderweb protocluster at $z=2.16$ \citep{2024A&A...692A..22Z}, though the Spiderweb has a lower degree of overdensity ($\sim2\times$). As also noted by \citet{2024A&A...692A..22Z}, the higher number of $>1$ mJy sources in SSA22 than Spiderweb could be due to our ALMA footprint covering only the very center of the $z = 3.09$ protocluster and an earlier evolutionary stage of the SSA22 protocluster when there was a large amount of gas supply from the cosmic web. In this context, the number count of the SSA22 field suggests that the early protocluster core environment preferentially enhances the massive and hence bright end of the star-forming population and that a potential dearth of faint DSFGs in the protocluster core, despite our survey being $90\%$ complete down to a 1.17 mm flux density of 0.19 mJy. The findings here agree with the galaxy downsizing scenario \citep[e.g.,][]{1996AJ....112..839C,2005ApJ...619L..43A}: more massive galaxies form earlier and faster, producing brighter dust emission. 

\subsection{The rest-frame FIR luminosity function}
In Fig. 4, we can see that the rest-frame 250 $\mu$m luminosity function of the SSA22 protocluster core region is $\sim2$ dex above that in coeval blank field \citep[e.g.,][]{2011ApJ...742...24L,2017MNRAS.471.4155K,2020A&A...643A...8G,2024ApJS..275...36F}. Such orders of magnitude higher obscured star formation rate densities than those in blank fields have also been observed in other high-redshift protoclusters \citep[e.g.,][]{2018ApJ...856...72O,2019MNRAS.488.1790L,2020MNRAS.495.3124H,2024A&A...692A..22Z}. In line with what has been seen in number count, the luminosity function shows evidence of a declining (or at least flat) faint end, suggesting a lack of faint DSFGs in the SSA22 protocluster core region: if the functional form is the same as blank fields, then by scaling the $z\sim3$ field luminosity function to match the two brightest flux bins of SSA22, we will expect much more sources at S$_{\rm 1.17mm}<1$ mJy than what has been observed. The overall shape of the rest-frame 250 $\mu$m luminosity function of the SSA22 protocluster core region is reminiscent of the Gaussian-shape bright end of the optical luminosity function of low-redshift galaxy clusters \citep[e.g.,][]{1997A&A...317..385L,1997MNRAS.284..915W,2003AJ....125.1849B}. This points to an evolutionary link between protocluster and local galaxy clusters and supports the idea that the massive early-type galaxies in local clusters have formed the bulk of their stellar mass during star formation events enshrouded by dust at $z\gtrsim2-3$ \citep[e.g.,][]{2017ApJ...849..154S,2019MNRAS.486.3047C,2024MNRAS.529.2290S}. Finally, it is notable that recent high-resolution JWST and ALMA observations of DSFGs in the SSA22 protocluster core have revealed evidence of growing bulge components behind dust obscuration \citep{2025arXiv250201868U}, which again suggests that we are witnessing the early formation phase of the massive end of the galaxy population in a protocluster.
\begin{figure}[!htp]
    \centering
    \includegraphics[width=0.99\linewidth]{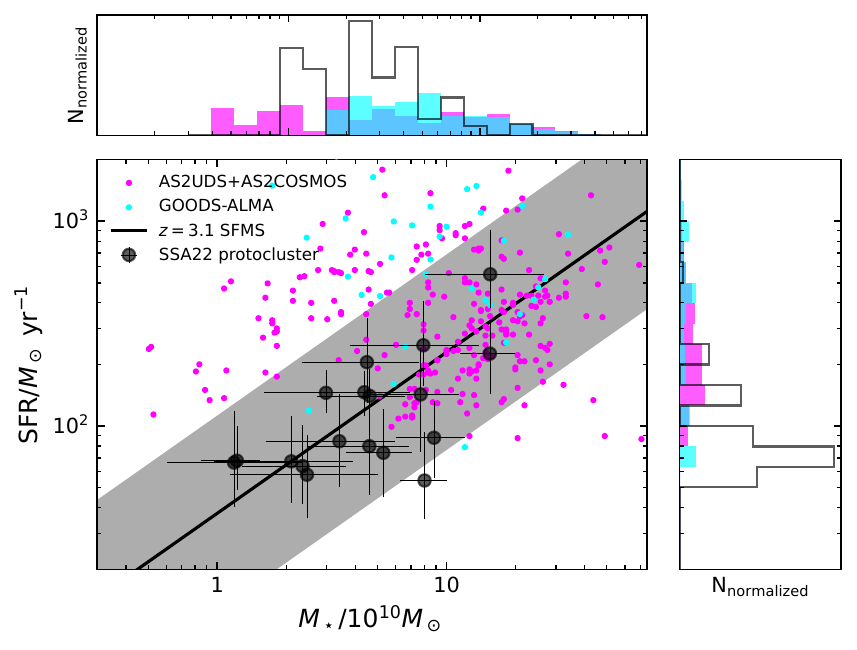}
    \caption{The SSA22 protocluster member galaxies (black circles) in the $M_\star$-SFR plane. The solid line shows the SFMS at $z=3.1$ from \citet{2014ApJS..214...15S}. The shaded regions indicate $3\times$ difference from the SFMS. The magenta and cyan circles show $2.8<z<3.4$ DSFGs from ALMA blank field mosaic \citep[GOODS-ALMA,][]{2018A&A...620A.152F,2020A&A...643A..30F} and single-dish follow-up \citep[AS2UDS+AS2COSMOS,][Ikarashi et al. in prep]{2020MNRAS.494.3828D,2020MNRAS.495.3409S} surveys.}
    \label{fig6}
\end{figure}
\subsection{Stellar mass and SFR of the SSA22 protocluster member galaxies}
In Fig. 5, the SSA22 protocluster member galaxies have SFR overlapping with the SFMS, but the distributions of $M_\star$ and SFR are different from field DSFGs. The differences are at $>3\sigma$ significance levels from a two-sided KS test except for the $M_\star$ distributions of SSA22 protocluster members and GOODS-ALMA DSFGs, which differ at $>2\sigma$. Overall, SSA22 protocluster galaxies have lower SFR and smaller $M_\star$, and we note that it is unlikely that the differences are driven by the choice of SED fitting codes because their results usually agree within error \citep{2023ApJ...944..141P}. Instead, it can be driven by the fainter fluxes of our sample: the median 1.17 mm fluxes are 0.8, 1.3 and 2.0 mJy for SSA22 protocluster member, GOODS-ALMA and AS2UDS+AS2COSMOS DSFGs, respectively. The deeper flux limit of our survey can pick up lower SFR and less massive systems, and the median $M_\star$ of each sample ($4.6\times10^{10}M_\odot$, $8.5\times10^{10}M_\odot$, $12\times10^{10}M_\odot$) roughly scales as the median 1.17 mm fluxes. Since the submm/mm fluxes reflects a combination of dust mass and SFR ($\sim M_{\rm dust}^{0.7}{\rm SFR}^{0.3}$), this is the expected behavior if we further assume $M_{\rm dust}\propto M_\star$ and consider the fact that none of the SSA22 protocluster member galaxies have SFR $3\times$ above SFMS like the field DSFGs. As $M_{\rm dust}$ is a proxy to total gas mass \citep[e.g.,][]{2016ApJ...820...83S,2020RSOS....700556H}, the above results suggest the current star formation in SSA22 protocluster member galaxies occurs in a self-regulated way powered by a large amount of fresh gas supply at the $z=3.09$ cosmic web node \citep{2019Sci...366...97U}, rather than short-lived starbursts induced by major merger. In addition, as will be presented in a future paper, [C \textsc{ii}] observations reveal that the majority of bright SSA22 protocluster member DSFGs are regular rotating disks with widespread star formation \citep[see also,][]{2025arXiv250201868U} instead of late-stage major merger, and the brightest one, A1, is a massive spiral galaxy \citep{2024arXiv241022155U}. These observations further support our argument about secular star formation. A similar conclusion has also been drawn for the Spiderweb protocluster \citep{2024ApJ...977...74P}.

\section{Summary}\label{section:con}
In this paper, we present a deep ALMA 1.17 mm continuum mosaic observation to study the faint DSFG population in an 8 arcmin$^2$  core region of the $z=3.09$ SSA22 protocluster. The continuum image has $\approx1\arcsec$ resolution and an RMS of $\approx25$ $\mu$Jy beam$^{-1}$, $\approx2.4\times$ of the depth in the previous study \cite{2017ApJ...835...98U}. Adopting an S/N$>4.2$ cut, we detect 53 continuum sources, doubling the number of detections in the same field. We search for counterparts at other wavelengths using optical-to-mid infrared imaging data of the SSA22 field. For 43 of 44 sources with robust counterparts in NIR bands, redshift, $M_\star$ and SFR are also derived from the photometric data. We identify 18 out of 53 sources as SSA22 protocluster members, with spectroscopic redshifts within $\Delta z=0.1$ from $z=3.09$.
\par
We compute 1.17 number count and find (1) an excess of 1.17 mm continuum sources at brighter fluxes ($\gtrsim2\times$ blank fields at $\gtrsim1$ mJy), which is produced by the $z=3.09$ SSA22 protocluster; (2) agreement with blank field surveys after excluding SSA22 protocluster member galaxies. The observed excess of bright 1.17 mm sources is driven by overabundance of massive dusty star-forming galaxies in the SSA22 protocluster core, but SFR in individual member galaxies is not significantly elevated compared to the expectation for their stellar masses, suggesting the SSA22 protocluster core environment enhance the formation of massive galaxies mainly by huge amount of gas supply from the cosmic web instead of frequent major mergers. We also compute the rest-frame 250 $\mu$m luminosity function of the protocluster core region and find evidence of a declining faint end, mimicking the shape of the bright end optical luminosity function of low-redshift galaxy clusters. These results suggest the ongoing build-up of a future galaxy cluster, adding to the growing body of evidence that massive early-type galaxies in local clusters form the bulk of their stellar mass in dusty star formation episodes at high redshift.
\begin{acknowledgements}
This work is supported by the NAOJ ALMA Scientific Research Grant Code 2024-26A and JSPS grant-in-aid No. 22H04939, 23K20035, 24H00004. IRS acknowledges support from STFC (ST/X001075/1). This paper makes use of the following ALMA data: ADS/JAO.ALMA \#2013.1.00162.S, \#2017.1.01332.S, \#2022.1.00680.S and \#2023.1.01206.S. ALMA is a partnership of ESO (representing its member states), NSF (USA), and NINS (Japan), together with NRC (Canada), MOST and ASIAA (Taiwan), and KASI (Republic of Korea), in cooperation with the Republic of Chile. The Joint ALMA Observatory is operated by ESO, AUI/NRAO and NAOJ. Some of the data presented in this paper were obtained from the Mikulski Archive for Space Telescopes (MAST) at the Space Telescope Science Institute. The specific observations analyzed can be accessed via \url{https://doi.org/10.17909/d5ma-0904}. STScI is operated by the Association of Universities for Research in Astronomy, Inc., under NASA contract NAS5–26555. Support to MAST for these data is provided by the NASA Office of Space Science via grant NAG5–7584 and by other grants and contracts.
\end{acknowledgements}
\bibliographystyle{aa_url}
\bibliography{main}\let\cleardoublepage\clearpage
\begin{appendix}
{\onecolumn
\section{Source catalogue}
\begin{table*}[hbtp!]
 \caption{\label{tablea1}List of S/N$>4.2$ sources detected in deep SSA22 1.17 mm mosaic.}{\centering\footnotesize
\begin{tabular}{lcclcccllcc}
 \hline \hline
  ID &RA &Dec &S/N &S$_{\rm 1.17mm}$ &NIR\tablefootmark{a} & redshift\tablefootmark{b}
  &spectral feature\tablefootmark{c} &Reference &$M_\star$ &SFR
 \\ \hline
\multicolumn{3}{l}{\tiny ID matches \cite{2017ApJ...835...98U}\tablefootmark{d}} &&[mJy]&&&&&[$10^{10}M_{\odot}$]&[$M_\odot$/yr]\\
\hline
A1 &22:17:32.42 &00:17:43.8 &178.1 &$6.149\pm0.060$ &Y &3.090 & CO(3\textendash2), [O \textsc{iii}]&(1)&$15.5_{-7.5}^{+11.2}$&$551_{-171}^{+351}$\\
A5 &22:17:31.48 &00:17:58.0 &81.4 &$2.530\pm0.055$ &Y &3.089 &CO(3\textendash2), [O \textsc{iii}]&(1)&$4.5_{-2.1}^{+3.1}$&$205_{-75}^{+131}$\\
A4 &22:17:36.96 &00:18:20.7 &80.2 &$2.212\pm0.028$ &Y &3.090 &CO(3\textendash2), CO(9\textendash8)&(2)&$7.9_{-4.1}^{+5.9}$&$248_{-91}^{+160}$\\
A3 &22:17:35.15 &00:15:37.2 &74.4 &$2.230\pm0.036$ &Y &3.096 &CO(3\textendash2), [CI](1\textendash0)&(3)&$3.0_{-1.4}^{+1.9}$&$145_{-29}^{+42}$\\
A2 &22:17:36.11 &00:17:36.7 &83.7 &$2.170\pm0.042$ &Y &3.991 &CO(4\textendash3), [CI](1\textendash0)&(4)&$12.3_{-9.8}^{+43.2}$&$198_{-97}^{+224}$\\
A7 &22:17:32.20 &00:17:35.6 &70.2 &$2.033\pm0.039$ &Y &3.094 &CO(3\textendash2)&(1)&$4.4_{-2.5}^{+2.6}$&$146_{-34}^{+39}$\\
A6 &22:17:35.84 &00:15:59.0 &54.5 &$1.427\pm0.042$ &Y &3.095 &CO(3\textendash2), [O \textsc{iii}]&(1), (5)&$4.6_{-1.9}^{+2.0}$&$140_{-40}^{+50}$\\
A8 &22:17:37.11 &00:17:12.3 &43.9 &$1.137\pm0.043$ &Y &3.090 &[O \textsc{iii}]&(5)&$8.8_{-2.8}^{+3.3}$&$87_{-31}^{+44}$\\
A11 &22:17:37.05 &00:18:22.2 &28.9 &$0.853\pm0.029$ &Y &3.091 &CO(3\textendash2), [O \textsc{iii}]&(5) &$7.7_{-3.3}^{+3.6}$&$142_{-67}^{+98}$\\
A10 &22:17:37.10 &00:18:26.8 &22.3 &$0.821\pm0.054$ &Y &3.091 &CO(3\textendash2)&(1)&$1.2_{-0.6}^{+1.2}$&$66_{-26}^{+52}$\\
A14 &22:17:31.33 &00:16:39.6 &24.1 &$0.805\pm0.040$ &Y &3.098 &CO(3\textendash2), [O \textsc{iii}]&(1)&$3.4_{-1.8}^{+2.5}$&$84_{-34}^{+58}$\\
A13 &22:17:37.43 &00:17:32.4 &23.5 &$0.706\pm0.039$ &Y &3.090 &CO(3\textendash2), [O \textsc{iii}]&(1)&$5.3_{-1.7}^{+1.8}$&$74_{-28}^{+46}$\\
A9 &22:17:36.54 &00:16:22.6 &29.1 &$0.689\pm0.025$ &Y &3.094 &CO(3\textendash2), [O \textsc{iii}]&(1), (5)&$15.5_{-4.0}^{+4.4}$&$226_{-81}^{+103}$\\
A17 &22:17:37.67 &00:18:14.6 &13.8 &$0.675\pm0.052$ &Y &3.091 &CO(3\textendash2)&(1)&$2.1_{-1.1}^{+1.8}$&$67_{-24}^{+44}$\\
A12 &22:17:32.01 &00:16:55.5 &20.6 &$0.573\pm0.030$ &Y &3.090 &[O \textsc{iii}]&(5)&$8.0_{-1.8}^{+2.0}$&$54_{-19}^{+39}$\\
A18 &22:17:32.24 &00:15:27.8 &10.6 &$0.562\pm0.061$ &Y &2.105 &Ly$\alpha$, N \textsc{v}, C \textsc{iv}, He \textsc{ii}&(6)&$4.1_{-1.8}^{+4.9}$&$148_{-54}^{+48}$\\
A16 &22:17:36.81 &00:18:18.0 &9.3 &$0.410\pm0.058$ &Y &3.087 &CO(3\textendash2), [O \textsc{iii}]&(1), (5)&$2.4_{-1.1}^{+1.3}$&$63_{-22}^{+36}$\\
A15 &22:17:32.76 &00:17:27.6 &12.0 &$0.326\pm0.036$ &Y &3.086 &CO(3\textendash2) &this work&$4.6_{-2.1}^{+2.3}$&$79_{-33}^{+48}$\\
A20 &22:17:34.65 &00:16:35.4 &8.3 &$0.305\pm0.037$ &Y &1.399 &CO(2\textendash1)&this work&$3.9_{-1.2}^{+1.0}$&$43_{-10}^{+45}$\\
A21 &22:17:33.08 &00:17:18.7 &6.5 &$0.267\pm0.039$ &Y &0.709 &[O \textsc{ii}], H$\beta$, [O \textsc{iii}]&this work&$5.3_{-1.2}^{+1.7}$&$46_{-19}^{+49}$\\
A28 &22:17:32.50 &00:17:29.4 &6.9 &$0.161\pm0.029$ &Y &0.641 &[O \textsc{ii}], H$\beta$, [O \textsc{iii}]&this work&$4.9_{-1.1}^{+2.0}$&$11_{-9}^{+20}$\\
\hline
\multicolumn{9}{l}{ New sources (${\rm S/N}\geq5.0$)} \\
\hline
A34 &22:17:31.12 &00:17:15.7 &14.4 &$0.424\pm0.040$ &Y &$2.93_{-0.66}^{+0.57}$ &...&...&$2.4_{-1.2}^{+1.8}$&$55_{-22}^{+39}$\\
A35 &22:17:30.55 &00:18:33.6 &5.1 &$0.366\pm0.076$ &Y &$2.67_{-0.20}^{+0.07}$ &...&...&$4.5_{-1.3}^{+0.9}$&$39_{-16}^{+20}$\\
A36 &22:17:38.63 &00:15:40.1 &6.4 &$0.360\pm0.096$ &Y &1.397 &CO(2\textendash1)&this work&$9.7_{-1.9}^{+1.6}$&$63_{-30}^{+37}$\\
A37 &22:17:29.74 &00:17:15.0 &6.4 &$0.349\pm0.073$ &Y &$2.08_{-0.02}^{+0.04}$ &...&...&$2.1_{-0.7}^{+0.9}$&$78_{-27}^{+50}$\\
A38 &22:17:30.52 &00:18:05.5 &12.5 &$0.372\pm0.038$ &Y &3.087 &CO(3\textendash2)&this work&$2.5_{-1.3}^{+2.6}$&$57_{-22}^{+32}$\\
A41 &22:17:31.76 &00:17:43.7 &6.9 &$0.298\pm0.044$ &Y &3.098&CO(3\textendash2)&this work&$1.2_{-0.4}^{+0.3}$&$67_{-19}^{+32}$\\
A42 &22:17:31.06 &00:15:38.5 &5.4 &$0.248\pm0.062$ &Y &$2.34_{-0.05}^{+0.07}$ &...&...&$2.0_{-0.5}^{+0.5}$&$24_{-10}^{+14}$\\
A43 &22:17:32.11 &00:17:28.4 &7.4 &$0.262\pm0.045$ &Y &0.641 &[O \textsc{ii}], H$\beta$, [O \textsc{iii}]&this work&$3.8_{-1.0}^{+0.8}$&$29_{-10}^{+15}$\\
A44 &22:17:38.05 &00:17:59.9 &8.7 &$0.259\pm0.025$ &Y &$1.72_{-0.07}^{+0.12}$ &...&...&$2.8_{-0.7}^{+1.1}$&$37_{-9}^{+18}$\\
A45 &22:17:33.75 &00:15:29.9 &5.8 &$0.234\pm0.061$ &Y &$2.35_{-0.12}^{+0.19}$ &...&...&$4.8_{-0.9}^{+0.6}$&$7_{-4}^{+8}$\\
A46 &22:17:34.59 &00:17:54.5 &8.6 &$0.241\pm0.041$ &Y &1.511 &Fe \textsc{ii}, Mg \textsc{ii}, Mg \textsc{i}&this work&$3.3_{-0.9}^{+1.1}$&$61_{-25}^{+46}$\\
A48 &22:17:32.68 &00:18:08.8 &5.0 &$0.191\pm0.041$ &Y &2.688 &C \textsc{ii}, C \textsc{iv}, C \textsc{iii}]&this work&$7.8_{-0.8}^{+0.9}$&$28_{-10}^{+14}$\\
A49 &22:17:30.74 &00:17:06.5 &8.0 &$0.209\pm0.043$ &Y &$2.67_{-0.38}^{+0.58}$ &...&...&$6.5_{-2.9}^{+3.5}$&$33_{-18}^{+31}$\\
A50 &22:17:37.45 &00:17:00.7 &5.3 &$0.187\pm0.036$ &Y &$3.92_{-0.13}^{+0.10}$ &...&...&$1.6_{-0.5}^{+0.6}$&$41_{-14}^{+22}$\\
A52 &22:17:30.50 &00:17:36.8 &6.8 &$0.195\pm0.035$ &Y &$2.28_{-0.03}^{+0.06}$ &...&...&$3.8_{-1.2}^{+0.7}$&$26_{-9}^{+12}$\\
A54 &22:17:37.58 &00:16:10.3 &5.0 &$0.175\pm0.036$ &Y &$1.53_{-0.03}^{+0.06}$ &...&...&$0.2_{-0.1}^{+0.1}$&$12_{-5}^{+5}$\\
A55 &22:17:34.80 &00:17:02.7 &5.8 &$0.158\pm0.031$ &Y &$3.71_{-1.36}^{+2.49}$ &...&...&$2.3_{-1.3}^{+1.7}$&$26_{-13}^{+24}$\\
\hline
\multicolumn{9}{l}{Tentative new sources ($4.2<{\rm S/N}<5.0$)} \\
\hline
A33 &22:17:38.83 &00:18:21.5 &4.5 &$0.498\pm0.140$ &L/N &...&...&...&...&...\\
A39 &22:17:34.63 &00:18:32.3 &4.9 &$0.314\pm0.079$ &Y &$0.47_{-0.19}^{+0.91}$ &...&...&$1.2_{-1.1}^{+11.6}$&$21_{-21}^{+41}$\\
A40 &22:17:29.34 &00:16:25.3 &4.4 &$0.289\pm0.079$ &N  &...&...&...&...&...\\
A47 &22:17:37.29 &00:16:39.2 &4.4 &$0.184\pm0.040$ &Y &$2.85_{-0.08}^{+0.13}$ &...&...&$2.3_{-0.4}^{+0.4}$&$19_{-6}^{+12}$\\
A51 &22:17:29.83 &00:15:55.7 &4.3 &$0.172\pm0.049$ &N  &...&...&...&...&...\\
A53 &22:17:33.80 &00:17:02.7 &4.2 &$0.166\pm0.059$ &Y &0.876 &[O \textsc{ii}]&this work&$2.6_{-0.6}^{+0.7}$&$24_{-8}^{+11}$\\
A56 &22:17:30.22 &00:18:02.0 &4.6 &$0.135\pm0.035$ &N  &...&...&...&...&...\\
A57 &22:17:35.00 &00:15:29.6 &4.8 &$0.135\pm0.037$ &N  &...&...&...&...&...\\
A58 &22:17:37.04 &00:18:02.0 &4.5 &$0.119\pm0.029$ &L/N  &...&...&...&...&...\\
A59 &22:17:31.58 &00:17:01.8 &4.3 &$0.116\pm0.027$ &N  &...&...&...&...&...\\
A60 &22:17:35.80 &00:17:28.8 &4.7 &$0.100\pm0.038$ &N &...&...&...&...&...\\
A61 &22:17:32.12 &00:18:23.9 &4.3 &$0.078\pm0.028$ &Y &$2.34_{-0.07}^{+0.09}$ &...&...&$2.2_{-0.6}^{+0.5}$&$18_{-7}^{+13}$\\
A62 &22:17:36.35 &00:17:50.1 &4.4 &$0.074\pm0.028$ &Y &$2.02_{-1.57}^{+0.22}$ &...&...&$0.6_{-0.3}^{+0.2}$&$11_{-5}^{+7}$\\
A63 &22:17:34.74 &00:16:20.1 &4.6 &$0.072\pm0.044$ &N  &...&...&...&...&...\\
A64 &22:17:31.56 &00:17:06.1 &4.3 &$0.065\pm0.017$ &L/N  &...&...&...&...&...\\
\hline
\end{tabular}}
\tablebib{(1)~\citet{2019Sci...366...97U};
(2) \citet{2017PASJ...69...45H}; (3) \cite{2013MNRAS.435.1493A};
(4) \citet{2020A&A...640L...8U};
(5) \citet{2014ApJ...795..165S}; (6) \citet{2004ApJ...614..671C}
}
\tablefoottext{a}{S/N$>3$ in any of the NIR $Ks$ or Spitzer IRAC1/2 bands, L indicates potential lensed source.}
\tablefoottext{b}{When a spectral line detection is not available, we report photometric redshifts from the EAZY code.}
\tablefoottext{c}{For SSA22 member galaxies with multiple spectral lines observed, we report the line(s) from which the redshift is measured.}
\tablefoottext{d}{A* is the abbreviation of ADF22.* in \citet{2017ApJ...835...98U}.}
\end{table*}}
\FloatBarrier
{\twocolumn
\section{Note on new spectroscopic redshifts}
\begin{figure}[!htp]
    \centering
    \includegraphics[width=0.99\linewidth]{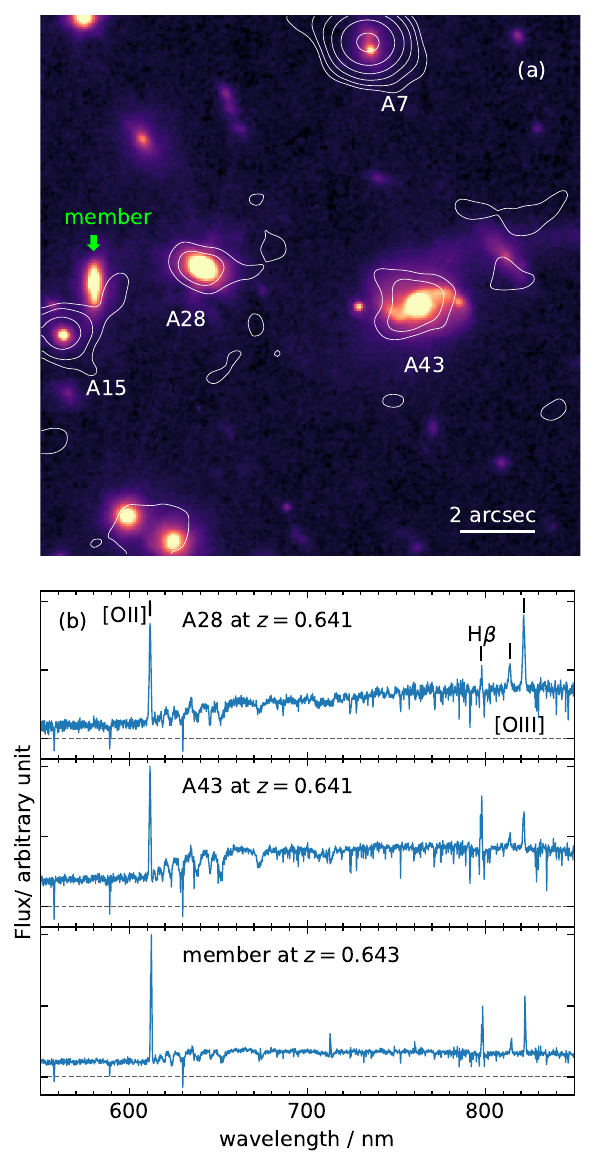}
    \caption{(a) JWST/NIRCam F444W image of the A28/A43 group. ALMA 1.17 mm continuum is overlaid as white contours in levels of 2, 4, 8, 16, 32...$\times$RMS. (b) VLT/MUSE spectra of A28, A43, and another group member. The horizontal dashed line indicates zero flux.}
    \label{figb1}
\end{figure}
\begin{figure}[!htp]
    \centering
    \includegraphics[width=0.99\linewidth]{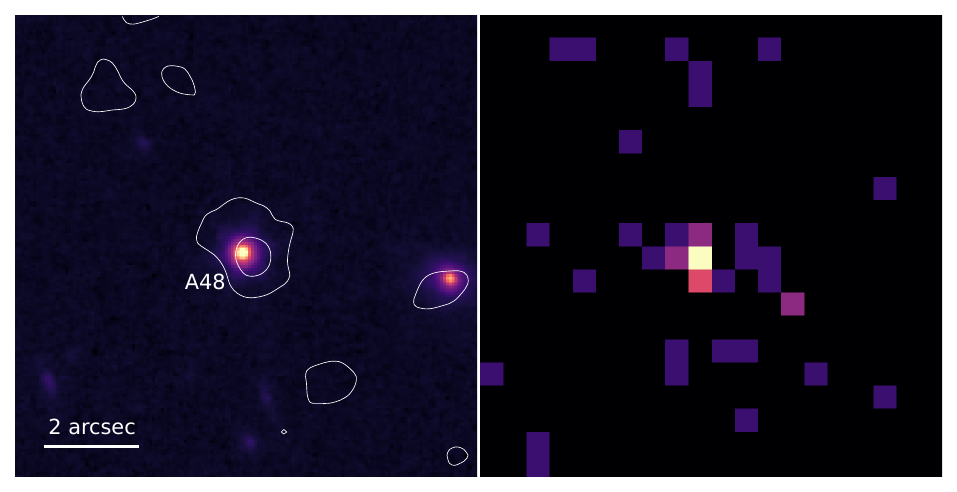}
    \caption{JWST/NIRCam F444W (left) and Chandra 0.5\textendash8 keV images of A48. ALMA 1.17 mm continuum is overlaid as white contours in levels of 2, 4$\times$RMS.}
    \label{figb2}
\end{figure}
\begin{figure*}
    \centering
    \includegraphics[width=0.99\linewidth]{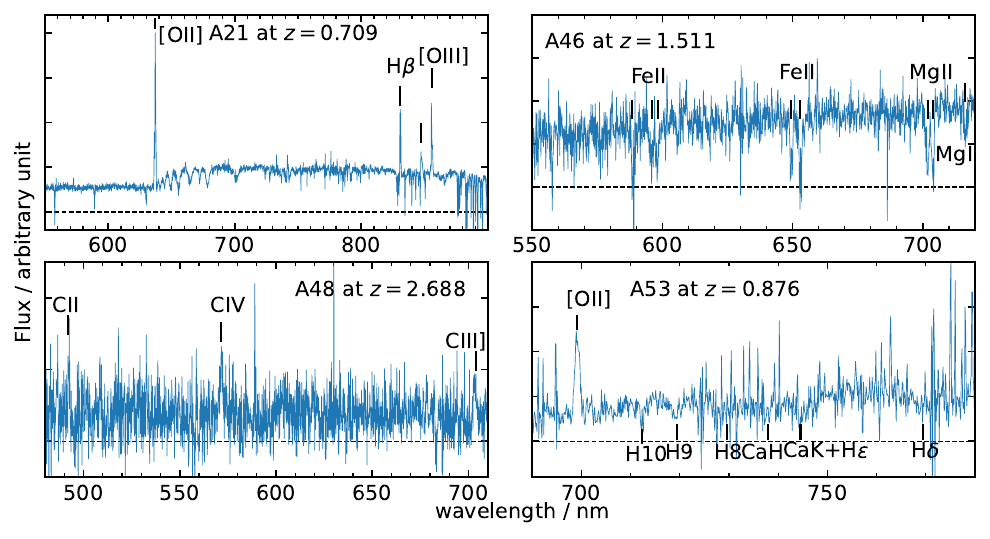}
    \caption{MUSE spectra of A21, A46, A48 and A53. The identified spectral features are marked.}
    \label{figb3}
\end{figure*}
\begin{figure*}
    \centering
    \includegraphics[width=0.99\linewidth]{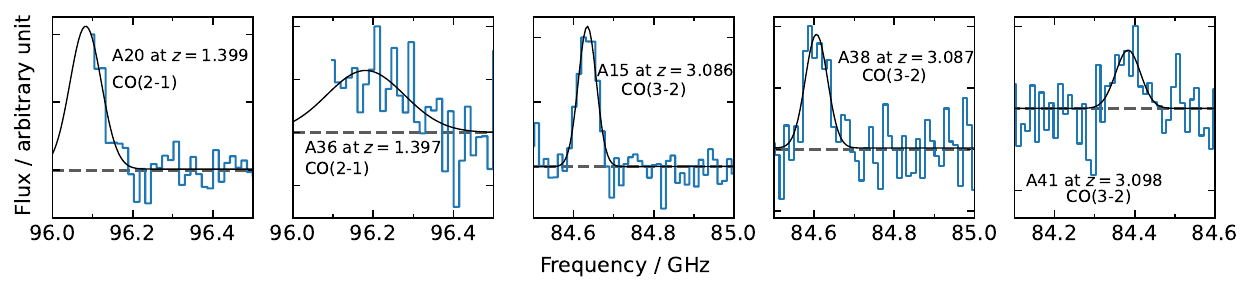}
    \caption{ALMA band 3 spectra of A20, A36, A15, A38 and A41. The black curves indicate the best-fit Gaussian profile of the CO lines.}
    \label{figb4}
\end{figure*}
In this section, we give a brief overview of new spectroscopic redshift measurements. For these sources, the VLT/MUSE spectra are shown in Fig. B.1, B.3 and ALMA band 3 spectra are shown in Fig. B.4. The sources and their spectral features are described below.
\\{\bf A28 and A43:} The detection of [O \textsc{ii}], H$\beta$ and [O \textsc{ iii}] emission lines suggests these two sources are members of a $z=0.64$ galaxy group, shown in panel (a) of Fig. B.1. Another member galaxy at $z=0.643$ is identified by the MUSE data but has no 1.1 mm counterpart (panel (b) of Fig. B.1). A tidal tail is visible on the north of A43. A28 is bright in the Spitzer MIPS 24 $\mu$m band, suggesting AGN activity.
\\ {\bf A21:} similar to A28 and A43, the source has multiple emission lines in the MUSE spectrum, which we identify as [O \textsc{ii}], H$\beta$ and [O \textsc{iii}] at $z=0.709$. The Balmer and metal absorption lines are also visible.
\\ {\bf A46:} There are many absorption features in the continuum, but no obvious emission line is detected. We identify the features to be Fe \textsc{ii}, Mg \textsc{ii} and Mg \textsc{i} absorption lines in rest-frame ultraviolet \citep{1993ApJS...86....5K} and determine its redshift $z=1.511$.
\\ {\bf A48:} We find three emission lines for this source: C \textsc{ii}, C \textsc{iv} and C \textsc{iii}] at $z=2.688$. The source hosts an AGN, as suggested by the broad C \textsc{iv} emission line (FWHM$\approx880$ km s$^{-1}$), detection in the Chandra X-ray data \citep{2009MNRAS.400..299L,2009ApJ...691..687L}, and a compact morphology in the JWST/NIRCam F444W image (Fig B.2).
\\ {\bf A53:} The source has only one emission line detected. We identify the line as [O \textsc{ii}] at $z=0.876$. Based on this redshift, we mark the expected position of strong absorption lines. H10, H9 and Ca \textsc{ii} K+H$\epsilon$ are clearly detected, but the remaining features are somewhat noisy.
\\ {\bf A20:} The source has one emission line on the edge of the ALMA Band 3 spectrum. The single-peaked photometric redshift of $z\sim1.5$ leads to a unique solution as CO(2\textendash1) at $z=1.399$.
\\ {\bf A36:} Similar to A20, CO(2\textendash1) at $z=1.397$ is assigned to the emission line on the edge of Band 3 spectrum based on its photometric redshift $z\sim1.4$.
\\ {\bf A15:} We update the redshift in \cite{2019Sci...366...97U} to $z=3.086$ using new CO(3\textendash2) data. It is close to the $z=0.643$ member galaxy of the A28/A43 group and has a point source-like appearance in JWST images and AGN signature. A15 is detected in Ly$\alpha$ emission.
\\ {\bf A38:} This source is identified as an SSA22 protocluster member, with redshift $z=3.087$ from the new CO(3\textendash2) detection.
\\ {\bf A41:} We update the [O \textsc{iii}] redshift in \cite{2019Sci...366...97U} to $z=3.098$ using new CO(3\textendash2) data. The source is a member of the SSA22 protocluster.
\FloatBarrier
\section{Postage stamp images}
\begin{figure*}[!htp]\centering\includegraphics[width=0.99\linewidth]{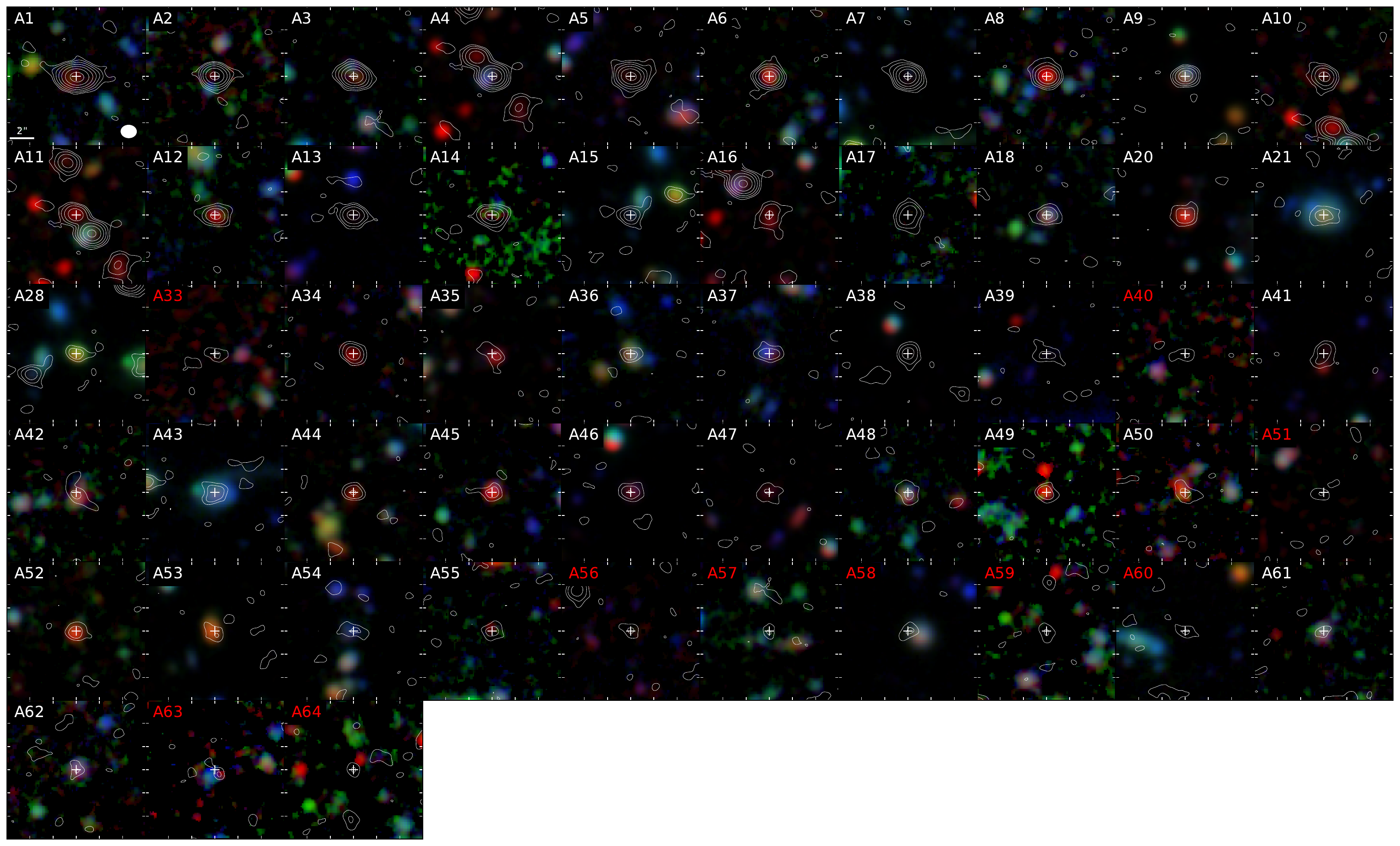}\caption{Ground-based imaging (R, G, B = $Ks$, $z^\prime$, $V$) of 1.17 mm sources (white contours in 2, 4, 8, 16...$\times$RMS levels). The images are smoothed to a common resolution of $\approx0\farcs8$. The ALMA beam is shown as the white ellipse. Each stamp is $12\arcsec\times12\arcsec$ in size. The source position is marked by the white cross. Sources without a reliable counterpart are marked as a red ID.}\label{fig:c1}\end{figure*}\begin{figure*}[!htp]\centering\includegraphics[width=0.99\linewidth]{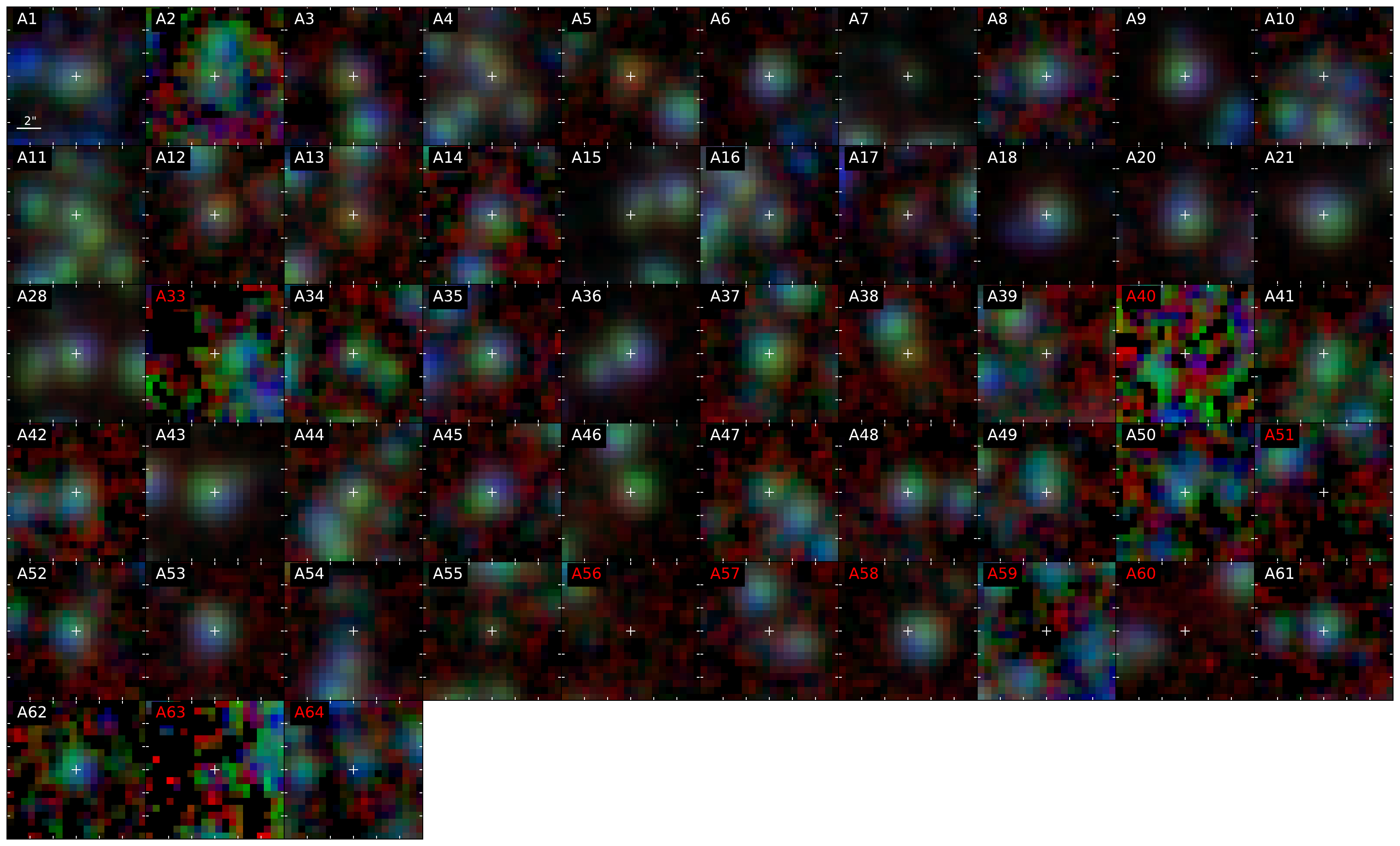}\caption{The same as Fig. C.1. but showing for Spizer IRAC data (R=IRAC3+IRAC4, G=IRAC2, B=IRAC1) withour 1.17 mm contour.}\label{fig:c2}\end{figure*}}
\end{appendix}
\end{document}